\documentclass[12pt,eqsecnum,preprint]{aastex}
\begin{document}
\title{PKS 1018$-$42: A Powerful Kinetically Dominated Quasar}
\author{Brian Punsly} \affil{4014 Emerald Street No.116,
Torrance CA, USA 90503 and International Center for Relativistic
Astrophysics, I.C.R.A.,University of Rome La Sapienza, I-00185
Roma, Italy} \email{brian.m.punsly@L-3com.com or
brian.punsly@gte.net}\and\author{Steven Tingay} \affil{Centre for
Astrophysics and Supercomputing, Swinburne
  University of Technology, P.O. Box 218, Hawthorn, Vic 3122, Australia}
\email{stingay@astro.swin.edu.au}

\begin{abstract}
We have identified PKS 1018-42 as a radio galaxy with extraordinarily 
powerful jets, over twice as powerful as any 3CR source of equal or lesser 
redshift except for one (3C196).  It is perhaps the most intrinsically 
powerful extragalactic radio source in the, still poorly explored, 
Southern Hemisphere.  PKS 1018-42 belongs to the class of FR II objects 
that are kinetically dominated, the jet kinetic luminosity, $Q \sim 6.5 \times
10^{46}\mathrm{ergs/s}$ (calculated at 151 MHz), is 3.4 times larger than the total 
thermal luminosity (IR to X-ray) of the accretion flow, $L_{bol} \sim 1.9 \times
10^{46}\mathrm{ergs/s}$.  It is the fourth most kinetically dominated quasar that we 
could verify from existing radio data.  From a review of the literature, 
we find that kinetically dominated sources such as PKS 1018-42 are rare, 
and list the 5 most kinetically dominated sources found from our review.  
Our results for PKS 1018-42 are based on new observations from the 
Australia Telescope Compact Array.  

\end{abstract}


\section{Introduction}The southern sky below a declination of
$-40^{\circ}$ is still not well explored in the radio band
compared to the northern sky and many of the intrinsically most
powerful radio sources might not have ever been imaged in detail at radio wavelengths. Suspecting
this to be the case, we began to search for evidence of the most
powerful Southern Hemisphere radio sources from archival spectral
information. Two sources stood out: PKS 0743$-$67 (reported in
\citet{pun06}); and PKS 1018$-$42.

In this Letter, we present the first deep radio observations of
PKS 1018$-$42, a quasar at $z=1.28$ \citep{hew93} with a 5 GHz
flux density over 1.2 Jy \citep{gre94}.  The steep radio spectrum
of PKS 1018$-$42 over a frequency range of 80 MHz \citep{sle95} to
31.4 GHz \citep{gel81} suggests that the source is dominated by
optically thin radio lobe emission rather than Doppler-boosted
core emission.  Due to its southerly declination, there are no
previously published deep radio maps of this extremely powerful
object, however \citet{ulv81} obtained VLA observations giving an
indication of some source structure, but were not
sufficient to reveal the detailed source morphology.  Basic structural information from the VLA observations was limited only
to the 20 cm waveband.

We have used Australia Telescope Compact Array (ATCA) observations to separate the core and lobe emission between
2.5 and 8.6 GHz, in order to caclulate the time-averaged kinetic luminosity of the jets, $Q\approx 6.5\times 10^{46}\mathrm{ergs/s}$,
making PKS 1018$-$42 one of the most kinetically dominated quasars known.

In this paper we adopt the following cosmological parameters: $H_{0}$=70 km/s/Mpc, $\Omega_{\Lambda}=0.7$
and $\Omega_{m}=0.3$.  We use the radio spectral index, $\alpha$, as $S\propto\nu^{-\alpha}$.

\section{The Radio Observations}
\par The Australia Telescope Compact Array (ATCA) was used to obtain
observations of PKS 1018$-$42 on December 16, 2001, in a series of 20 minute $``$cuts$''$ over a 12
hour period. For half of the cuts the ATCA was configured to
observe at frequencies of 1384 and 2496 MHz (1.4 and 2.5 GHz) and at 4800 and 8640 MHz (4.8 and 8.6 GHz) for the other half.  The observations
took place while the array was in a 6 km baseline configuration, allowing maximum angular resolution.  At all
frequencies the bandwidth was 128 MHz in each of two crossed linear polarisations.

Standard data reduction and imaging techniques were used to produce images from the data \citep{sau95}.
Stokes I, Q, and U images were produced from the 2.5, 4.8, and 8.6 GHz data.  It was found
that the 1.4 GHz data proved of insufficient angular resolution to be useful.  Figure 1 shows the resulting
image of PKS 1018$-$42 at 4.8 GHz.  A core is centrally located between two powerful
lobes. Estimates of component sizes and positions were made by model-fitting the \emph{u-v}
data with point sources, circular Gaussian, and elliptical Gaussian
components in the DIFMAP package \citep{she94} and are summarized in Table 1.

We are confident that we have recovered the full flux density of the source at each frequency since at 8.6 GHz,
our highest frequency and therefore highest angular resolution, our total measured flux density of 0.63 Jy matches
the independently determined single dish flux density at this frequency \citep{wri91}.  We note that the 8.6 GHz image implies
an axial length of 14.9\arcsec. In
our adopted cosmology, this corresponds to an axial length of
$D=140\mathrm{kpc}$.

\begin{figure}
\plotone{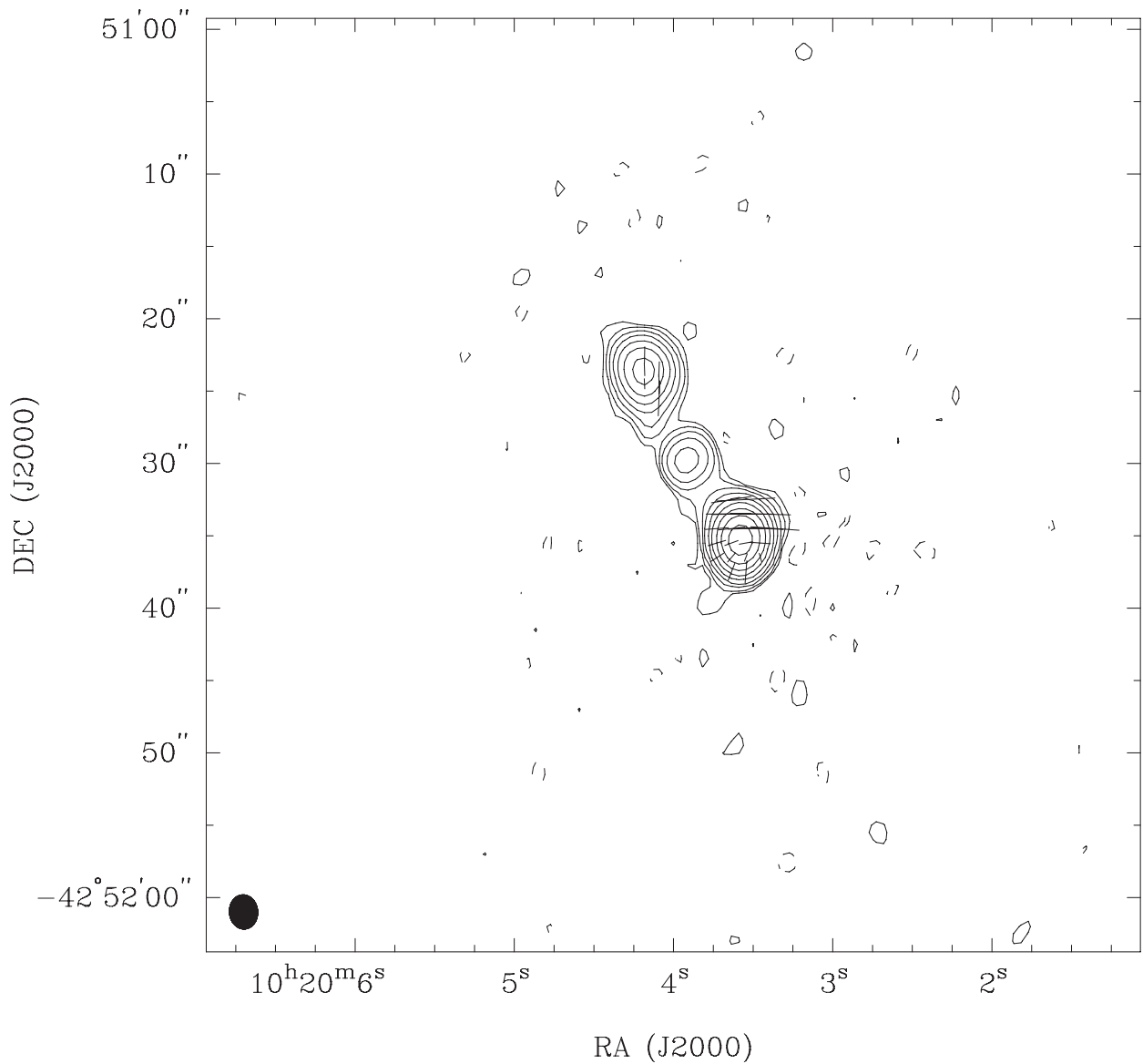}
    \caption{ PKS 1018$-$42 at 4.8 GHz.  The peak intensity in the
      image is 0.77 Jy/beam.  The beam-size is 1.9\arcsec$\times$2.3\arcsec at a
      position angle of $5.1^{\circ}$.  Contour levels for the Stokes
      I emission are 0.0077 Jy/beam $\times$ (-0.25, 0.25, 0.5, 1, 2,
      4, 8, 16, 32, 64).  The peak fractional polarization is 18.5\%. The vector lengths
      represent 4.4\% fractional polarization per arcsecond.}\end{figure}

\begin{table}
\caption{ATCA radio data for core and lobes of PKS 1018$-$42}
\begin{tabular}{cccccc}
\tableline \rule{0mm}{3mm}
Frequency &   Beam           &   Component& Flux     & FWHM            &   Notes\\
   MHz    &  arcsec          &            &   Jy     & arcsec          &        \\
\tableline \rule{0mm}{3mm}
1384      & 7.4 $\times$ 5.7 &  North     &  1.07    & 3.4 $\times$ 1.5&        \\
          &                  &  Core      &  na      & na              &       a\\
          &                  &  South     &  3.16    & 1.9 $\times$ 1.3&        \\
          &                  &  Total     &  4.23    &                 &        \\
          &                  &            &          &                 &        \\
2496      & 4.2 $\times$ 3.5 &  North     &  0.51    & 1.7 $\times$ 1.4&        \\
          &                  &  Core      &  0.12    & 5.7 $\times$ 1.3&       b\\
          &                  &  South     &  1.77    & 1.2 $\times$ 0.7&        \\
          &                  &  Total     &  2.40    &                 &        \\
          &                  &            &          &                 &        \\
4800      & 2.3 $\times$ 1.9 &   North    &  0.24    & 1.5 $\times$ 1.1&        \\
          &                  &   Core     &  0.06    & 1.9 $\times$ 0.0&       c\\
          &                  &   South    &  0.96    & 1.3 $\times$ 0.7&        \\
          &                  &   Total    &  1.26    &                 &        \\
          &                  &            &          &                 &        \\
8640      & 1.3 $\times$ 1.1 &   North    &  0.11    & 1.4 $\times$ 1.2&        \\
          &                  &   Core     &  0.03    & 0.5 $\times$ 0.0&        \\
          &                  &   South    &  0.49    &  na             &       d\\
          &                  &   Total    &  0.63    &                 &        \\
\tableline{\rule{0mm}{3mm}}
\end{tabular}\\
\tablenotetext{a}{Core not detected/resolved at 1384 MHz.}
\tablenotetext{b}{Core detected but appears highly extended in
direction of lobes.  Probably lobe emission contaminating estimate
of core flux and size.} \tablenotetext{c}{Core unresolved in
direction perpendicular to the lobe direction.}
\tablenotetext{d}{South lobe resolved into 4 sub components,
between 0.01 and 0.20 Jy in flux and 0.5 and 1.0 arcseconds in
size.  Flux given is sum of subcomponent fluxes.}
\end{table}

\section{Estimating the Jet Kinetic Luminosity}
We estimate the jet kinetic luminosity from the isotropic extended
emission, applying a method that allows one to convert 151 MHz
flux densities, $F_{151}$ (measured in Jy), into estimates of
kinetic luminosity, $Q$ (measured in ergs/s), following \citet{wil99} and \citet{blu00}, by means of the
formula derived in \citet{pun05}:
\begin{eqnarray}
&& Q \approx 1.1\times
10^{45}\left[(1+z)^{1+\alpha}Z^{2}F_{151}\right]^{\frac{6}{7}}\mathrm{ergs/sec}\;,\\
&& Z \equiv 3.31-(3.65) \nonumber \\
&&\times\left(\left[(1+z)^{4}-0.203(1+z)^{3}+0.749(1+z)^{2}
+0.444(1+z)+0.205\right]^{-0.125}\right)\;,
\end{eqnarray}
where $F_{151}$ is the total optically thin flux density from the
lobes (i.e., no contribution from Doppler boosted jets or radio
cores). The formula is most accurate for large classical double
radio sources which is the case for PKS 1018$-$42 \citep{wil99}.

A measurement of the lobe flux at low frequency is critical to the
successful implementation of (3.1).  Our estimate of the spectral
index between 2.5 and 8.6 GHz for the northern lobe is
$\alpha=1.23$ and for the southern lobe is $\alpha=1.03$.  We
estimate the core spectral index between 2.5 and 8.6 GHz as
$\alpha=1.14$.  We cannot separate the core and lobes at 1.4 GHz
due to poor angular resolution at this frequency.  However,
extrapolating the lobe spectra to 1.4 GHz and subtracting the lobe
fluxes at this frequency gives us an estimate of the core flux
density of 0.016 Jy.  The core spectrum therefore appears to
flatten to $\alpha=0.49$ near 1 GHz.  This being the case, the
core appearing to turn over below 1 GHz, the core contribution to
the total flux at 151 MHz will be negligible.

Extrapolating the lobe flux densities using the 2.5 to 8.6 GHz
spectral indices yields flux densities of 16 Jy and 45 Jy at 408
MHz and 160 MHz, respectively. However, the 408 MHz and 160 MHz
flux densities have been previously measured as 12.72 Jy,
\citet{lar81}, and 25.5 Jy, \citet{sle95}, respectively. Thus, the
lobe spectrum appears to flatten slightly at frequencies below 1
GHz. Using a spectral index derived from the measurements at 408
MHz and 160 MHz, and extrapolating to 151 MHz, yields
$F_{151}=26.4$, which can be used in (3.1) to give
$Q\approx6.5\times 10^{46}\mathrm{ergs/s}$. Alternatively, we can
use our new 4.8 GHz flux from Table 1 with equation (3.9) of
\citet{pun05}, adapted to 4.8 GHz flux densities, $F_{4800}$,
\begin{eqnarray}
&&Q \approx
1.81\times10^{46}(1+z)^{1+\alpha}Z^{2}F_{4800}\,\mathrm{ergs/sec}\;,\quad\alpha\approx
1\;.
\end{eqnarray}
to obtain a slightly larger estimate, $Q\approx9.5\times
10^{46}\mathrm{ergs/s}$.
\begin{table}
\caption{The Composite SED of a Radio Quiet Quasar $M_{V}=-25$}
\begin{tabular}{cccccc}
\tableline \rule{0mm}{3mm}
Log $\nu$ (Hz) & Log $\nu$ (Hz)  & $\nu F_{\nu}$& $\nu F_{\nu}$ & Band &  ref\\
  start &  end  & start & end  &  &\\
\tableline \rule{0mm}{3mm}
12.5 & 13.35 & 44.65  & 45.3 & mid IR & a\\
13.35 &  14.4 & 45.3  & 44.95 & near IR & a\\
14.4& 15.0 & 44.95 & 45.45 & optical & a,b,c\\
15.0& 15.4 & 45.45 & 45.6 & UV & b,c\\
15.4& 17.25& 45.6 & 44.3 & EUV/soft X-ray & b,c,d\\
17.25& 19.4 & 44.3 & 44.3 & X-ray &  d\\
\tableline{\rule{0mm}{3mm}}
\end{tabular}\\
\tablenotetext{a}{\citet{elv94}} \tablenotetext{b}{\citet{zhe97}}
\tablenotetext{c}{\citet{tel02}} \tablenotetext{d}{\citet{lao97}}
\end{table}

\section{Kinetic Dominance of the Source}
In this section we
estimate $R$, the ratio of the kinetic luminosity, $Q$, to the total bolometric luminosity
of the accretion flow, $L_{bol}$. Recall that $L_{bol}$ is the
bolometric luminosity of the thermal emission from the accretion
flow, including any radiation in broad
emission lines from photo-ionized gas or as IR reprocessed by
molecular gas. In order to estimate $L_{bol}$, we construct a
composite spectral energy distribution (SED) of a quasar accretion
flow \citep{pun06}. In order to separate the accretion flow thermal luminosity
from IR and optical contamination from the jet, an SED for radio
quiet quasars (normalized to $M_{V}=-25$) was chosen since this
represents pure accretion luminosity. A piecewise collection of
power laws in table 2 is used to approximate the individual bands
in the SED. The first two columns are the start and stop
frequencies for the each local power law. The third and fourth
columns are the corresponding values of $\nu F_{\nu}$. One can
compute the $L_{bol}$ of the accretion flow from the composite
SED. In addition to the continuum, the
      composite spectrum of \citet{zhe97} indicates that $\approx 25\%$ of the total optical/UV quasar
      luminosity is reprocessed in the broad line region.
      Combining this with the continuum luminosity
      yields $L_{bol}=1.35\times 10^{46}\mathrm{ergs/sec}$.
      Now if one assumes that the shape of the SED is unchanged with the magnitude of
      $L_{bol}$ then Table 2 is particularly useful, since it allows the
      knowledge of IR, UV or X-ray flux to estimate $L_{bol}$. Namely,
      $L_{bol}/ 1.35\times 10^{46}\mathrm{ergs/sec}$ scales
      with the value of the measured $\nu F_{\nu}$ divided by the
      value of $\nu F_{\nu}$ in the composite of Table 2 at the
      selected frequency of observation.
      With this assumption, the UV continuum flux density at the rest frame frequency of
      $1.37\times 10^{15}\mathrm{Hz}$ in the spectrum of PKS
      1018$-$42 in \citet{sti94} applied to the composite in Table 2 yields
      $L_{bol}=1.9\times 10^{46}\mathrm{ergs/s}$. Thus, $R=3.4$ for the 151 MHz estimate of $Q$ and
      $R=5$ for the 4.8 GHz estimate of Q.

\par It is interesting to ask how rare it is to find a
      kinetically dominated quasar, $R>1$. The natural place to
      find $R>1$ sources is to look at the highest redshift
      sources in a low radio frequency selected survey. To estimate $R$ in
      quasars, one can use primarily UV emission and also reprocessed IR dust emission to
      estimate $L_{bol}$ from Table 2 as a check
      for obscuration of the accretion disk. For FR II narrow line
      radio galaxies (hidden quasars), one can only use the
      reprocessed IR dust emission that peaks around $100 \mu m$.
      In addition, a high resolution radio map is required to
      isolate the optically thin radio flux density that is required
      in (3.1).
      \begin{table}
\caption{The Most Kinetically Dominated Quasars}
\begin{tabular}{ccccccc}
\tableline \rule{0mm}{3mm}
Source &  $z$ & $Q$& $R$  & freq & $\alpha $   &  ref\\
       &    & $10^{45}\mathrm{ergs/s}$  &  & $\nu(10^{15} Hz)$ & &\\
\tableline \rule{0mm}{3mm}
3C 82      & 2.878 &  155.4  &  10.7    & 0.014&0.7 & a\\
          &  2.878 &  155.4  &  6.22    & 1.67 &0.7 & a\\
3C 9      &  2.009 &  148.3  &  5.93    & 1.67& 0.36 & b\\
          &  2.009 &  148.3  &  3.82    & 0.0078& 0.36 & c\\
4C 45.21  & 2.686 &    59.3  &  5.11    & 1.14 &$-0.12\pm 0.3$ & b\\
PKS 1018-42 & 1.28 &   65.2   & 3.38    & 1.37 & 1.75 &  d\\
            & 1.28 &   65.2   & 4.45    & 1.07 & 1.75 &  d\\
3C 190      & 1.195 &  42.63  &  13.1   &  1.34 & $3.0\pm 0.5$& e\\
            & 1.195 &  42.63  &  1.12   &  0.133 & $3.0\pm 0.5$& f\\
            & 1.195 &  42.63  &   2.81  &  1.07 & $3.0\pm 0.5$& e\\
TXS 1243+036 & 3.57 &  114.8 &   2.71   & 0.0016& ... & g\\
3C 405       & 0.056 &  23.2 &   0.75   &  0.005& ... & h\\
             & 0.056 &  23.2 &   0.93   &  100 & ... & i\\
\tableline{\rule{0mm}{3mm}}
\end{tabular}\\
\tablenotetext{a}{\citet{sem04}} \tablenotetext{b}{\citet{bar90}
with data recalibrated by M. Vestergaard, private communication}
\tablenotetext{c}{\citet{mei01}} \tablenotetext{d}{\citet{sti94}}
\tablenotetext{e}{\citet{smi80}}\tablenotetext{f}{\citet{sim00}}
\tablenotetext{g}{\citet{arc01}\tablenotetext{h}{\citet{haa04}}}
\tablenotetext{i}{\citet{you02}}
\end{table}

\par To determine the rate of occurrence of kinetically dominated quasars,
the UV data from \citet{ver01} were cross-correlated with archival
VLA and MERLIN radio maps, in order to single out possible large
$R$ quasar candidates. The actual spectra (referenced in column 7
of table 3) of sources that looked promising were subsequently
studied explicitly in order to compute $L_{bol}$. Similarly, submm
data for high redshift, $z>3.0$, sources (redshifted dust
emission) in \citet{arc01} and IR data for $z<1.5$ sources from
\citet{mei01} and \citet{haa04} were used to estimate $L_{bol}$
for obscured quasars with powerful radio emission. An exhaustive
search of the literature revealed very few sources with $R$ larger
than that of PKS 1018$-$42. We found deep radio observations of $>800$
radio loud quasars primarily in the references
\citet{aku91,aku95,ant85,bog94,hin83,hut88,liu92,lon93,man92,mur93,nef89,nef90,pun95,rei99}
with $M_{V}$ tabulated in \citet{ver01}. Additionally, there were
another $\sim 100$ powerful FRII radio galaxies with IR data in
\citet{arc01,mei01,haa04} and deep radio observations. The large
$R$ sources are tabulated in Table 3. The first two columns are
the source and its redshift, followed by $Q$ and $R$ computed with
equation (3.1). The fifth column is the frequency in which a rest
frame flux density was used to estimate $L_{bol}$ from Table 2.
The sixth column is the UV spectral index, when it was available.
In spite of this search, two of the four highest $R$ sources in
Table 3, 3C 82 and PKS 1018$-$42 had no previously published radio
maps and were found by looking for high redshift sources in
\citet{ver91} with the largest low frequency flux densities and
steep spectral indices. These were the most promising candidates
out of the 6225 quasars. This initiated followup observations
reported here and in \citet{sem04}.
\par The two estimates for both 3C
82 (the highest redshift 3C quasar known) and 3C 9 (the highest
redshift 3CR quasar) in the IR and UV are in close agreement,
verifying the validity of the composite in Table 2. Both 3C 405
and 3C 190 were added to the list of the five largest $R$ sources
for comparison purposes.  3C 405 (Cygnus A) is the best studied
example of a very high $Q$ radio galaxy and the close agreement
between the X-ray and IR estimates for 3C 405 are also affirmation
of the application of Table 2. Note the difference in the
estimates for 3C 190. The IR and the continuum UV estimate at
$1.34\times 10^{15}\mathrm{Hz}$ disagree tremendously. This is
because 3C 190 is an obscured quasar with a red spectrum,
indicated by the steep UV spectral index of 3. Thus, a third
estimate for 3C 190 is given, $L_{bol} = 1.07\times
10^{15}\mathrm{Hz}$, from the MgII broad emission line. According
to \citet{wan04}, $L_{bol}\approx 165 L_{MgII}$, where $L_{MgII}$
is the line luminosity. This 3C 190 line estimate agrees with the
IR estimate, consistent with an attenuated line of sight toward
the accretion disk but not towards the low ionization broad line
region. The main reason for including 3C 190 is the rather steep
UV spectral index of PKS 1018$-$42. This might raise some concerns
about a heavily obscured accretion flow that is skewing the
estimate of $R$. We note that the MgII line estimator agrees with
the UV continuum estimate for PKS 1018-42. Even though this is
strongly supportive of the UV estimate of $R$, a space based IR
measurement would be definitive.

\section{Conclusion} In this Letter, we have investigated radio images of a very powerful
quasar PKS 1018$-$42 that is not well known due to its southerly
declination. However, PKS 1018-42 is extraordinarily powerful with
jets over twice as powerful as any 3CR source of equal or lesser
redshift except for one (3C196) and is certainly worthy of more
detailed study. We found that it is kinetically dominated with
$R\sim 3-5$. It is the fourth most kinetically dominated quasar
that we could verify from existing radio maps.
\par We showed in
Table 3 that kinetic dominance is a particularly rare circumstance
for quasars. This is consistent with the study of 7C sources in
\citet{wil99} that was based on estimating $L_{bol}$ from OII
narrow line emission. The OII narrow lines are very distant from
the central quasar and it is not clear how much jet propagation
excites narrow line emission (see \citet{vei97} for one of many
examples of narrow line emission that is stimulated by jet
propagation). Using the UV luminosity from the central quasar
directly as we have done is far more reliable, especially when
verified independently with IR and broad line data as was done
above. Thus, our results are consistent with \citet{wil99}, but
they are found from an independent and perhaps more scientifically
justified method. However, Table 3 is incomplete because the most
likely large $R$ candidates are high redshift 3C narrow line
galaxies and the current $450\mu m$ sensitivity required for these
estimates is not available, just loose bounds on the flux density
\citep{arc01}. Even if there are some kinetically dominated high z
radio galaxies, this should not change the conclusion drawn from
Table 3 that large values of $R$, such as for PKS 1018$-$42, are
unusual for quasars.

\par The rarity of large $R$ sources
follows from the steep luminosity distributions for FR II quasars,
$2\times 10^{45}\mathrm{ergs/s}< L_{bol}< 3\times
10^{48}\mathrm{ergs/s}$ and $10^{44}\mathrm{ergs/s}< Q < 3\times
10^{47}\mathrm{ergs/s}$ \citep{ver01,pun01}. A value of $R=10$
requires $Q > 2\times 10^{46}\mathrm{ergs/s}$ which is already at
the high end of the steep kinetic luminosity distribution and
these sources are rare. Yet, the complete absence of $R\sim 30$
sources still seems unexpected because $Q$ is a measure of the
long term activity of the radio source over $\sim 10^{7}$ years
and so it is not contemporaneous with the thermal emission from
the accretion flow \citep{pun05}. The timescale for the extended
emission is so long, it would seem that there must be sources in
which the accretion engine has long since shutoff in spite of
powerful radio lobe emission 100 kpc away. Where are the
$``$fossil$''$ sources with $Q \sim 5\times
10^{46}\mathrm{ergs/s}$ and the accretion flow almost shutoff,
$L_{bol}\sim 10^{45}\mathrm{ergs/s}$? Perhaps some can be found as
distant, $z>1.5$, 3C narrow line galaxies with improved IR
sensitivity. The Spitzer IR telescope is ideal for looking for
such hidden kinetically dominated quasars. PKS 1018-42 would be an
interesting source for the Spitzer telescope, in order to verify
its standing in table 3. However, Spitzer observations to date
have not added any new sources to table 3 \citep{haa05}.
Alternatively, the absence of such sources might tell us something
about the fundamental nature of the quasar jet central engine.

\end{document}